# Increasing single-photon production with cavity-QED

C. Shang,[1,2] W. W. Chow,[3] G. Moody[1,2], and J. E. Bowers[1,2]

[1] Department of Electrical and Computer Engineering, University of California, Santa Barbara, CA 93106 USA.

[2] Institute for Energy Efficiency, University of California, Santa Barbara, CA 93106 USA.

[3] Sandia National Laboratories, Albuquerque, NM 87185-1086, U.S.A.

*Abstract* — A study was performed to determine the extent cavity enhancement may increase single-photon production while maintaining single-photon purity. It was found that certain combinations of cavity lifetime and light-matter coupling strength can lead to carrier-photon correlations that increase single-photon generation rate while maintaining low 2$^{nd}$ order photon correlation. This study provides guidance for future emitter-cavity design to achieve high purity and ultrafast single photon generation.

## I. Introduction

The development of photonics-based quantum technologies is poised to revolutionize data processing and communication methods [1], enabling the on-demand generation and control of photonic qubits within established quantum networks [2]. To fully unlock the potential of these technologies, it is essential to have high-efficiency photon detectors, both linear and nonlinear photonic circuits, as well as reliable sources of single photons and entangled photon pairs. Despite the significant improvements that have been achieved for photonic circuits [3], [4], [5] and on-chip detectors [6], [7], [8], [9], the development of single photon sources has been slow.

Various physical systems have been explored for single-photon generation, such as semiconductor defects [10], [11], [12], [13], isolated ions [14], [15], spontaneous parametric down-conversion and four-wave mixing of laser pulses in nonlinear crystals[16], [17], [18], [19], [20], and semiconductor quantum dots (QDs) [21], [22], [23]. Though there is no clear leading deterministic single-photon source yet, semiconductor QDs are ideal two-level systems that could generate single photons on-demand via either optical [24] or electrical triggers [25]. Photons generated from various charge states can be harvested for different quantum applications, from memory-based quantum repeaters to cluster state generation for measurement-based quantum computing[26]. Epitaxially grown QDs, being a solid-state platform, also offers tunable photon properties and promises simple integration with on-chip quantum photonic architectures. Embedding QD emitters in optical cavities will further improve the spontaneous photon emission rate and directionality. A QD embedded in a micropillar still demonstrates record single photon generate rates across all material platforms. However, high single photon emission rates due to cavity enhancement generally poses a fundamental trade-off, whereby an increase in the photon emission rate degrades the single photon purity. In this work, utilizing cavity quantum electrodynamic (cQED), we investigate the possibility of locating a delicate balance between cavity lifetime and spontaneous emission rate to achieve higher single photon generation rate while maintaining high single photon purity. The challenge is maximizing cavity enhancement while ensuring no possibility of simultaneously having a residual photon in the cavity together with an excited emitter.

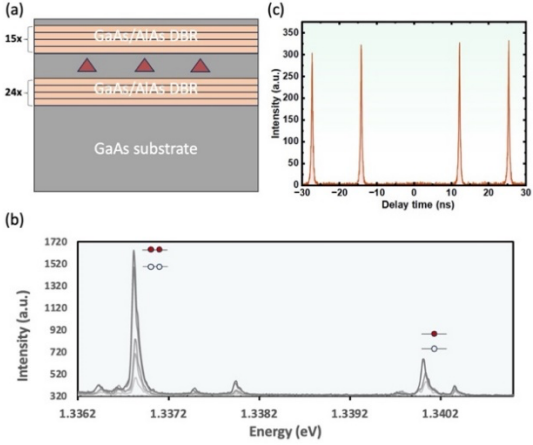

Figure 1: (a) Schematic of the material system under study. (b) Emission spectra of the QD under investigation with various pump power, from 0 to 512 µW. Emission peaks from the neutral exciton and biexciton are identified. (c) Autocorrelation measurement of the exciton emission under pulsed excitation.

## II. RESULTS AND DISCUSSION

The material system under investigation is a single InAs QD embedded in a distributed Bragg reflector (DBR) cavity grown by molecular beam epitaxy. The DBR cavity consists of a GaAs/AlAs superlattice as the top and bottom mirrors with a cavity lifetime of 0.3 ps and a cavity linewidth of approximately 1.5 nm. The measured second-order autocorrelation of the single photons emitted from a neutral exciton (X) is $g^{(2)}(0) = 0.008$. The detailed description of the material system and emission spectrum for the single QD can be found in [27]. Representative measurements of the QD under investigation are shown in Figure 1(a)-(c).

To investigate the possibility of maintaining or even improving the single photon purity at a higher photon generation rate, we examine the numerator and denominator of the second-order autocorrelation function at zero delay,

$$g^{(2)}(0) = \frac{\langle a^\dagger a^\dagger aa \rangle}{\langle a^\dagger a \rangle \langle a^\dagger a \rangle} , \qquad (1)$$

where $a^\dagger$ and $a$ are the photon creation and annihilation operators. In Equation 1, the denominator, $\langle a^\dagger a \rangle \langle a^\dagger a \rangle$, represents the square of the average intracavity photon number. It is the expected value of two independent photon detections, as would occur in a coherent or classical light field. The numerator, $\langle a^\dagger a^\dagger aa \rangle$, represents the true joint probability of detecting two or more photons simultaneously. Figure 2(a) and (b) show the calculated behavior of the intracavity photon number (denominator of $g^{(2)}(0)$) and the multiphoton probability (numerator of $g^{(2)}(0)$), respectively. In Figure 2(b), the difference in the cavity decay rate $\gamma_{cav} = 0.4\ ps^{-1}$ curve from the others explains the onset of desirable cQED contribution. The curve indicates a smaller increase in multiple photon probability $\langle a^\dagger a^\dagger aa\rangle$ with cavity lifetime compared to the others. Since $g^{(2)}(0)$ is essentially the normalized two photon probability with respect to uncorrelated photons, the smaller increase in the rate in the multiphoton probability with respect to the average intracavity photon number would result in "dips" in $g^{(2)}(0)$ values.

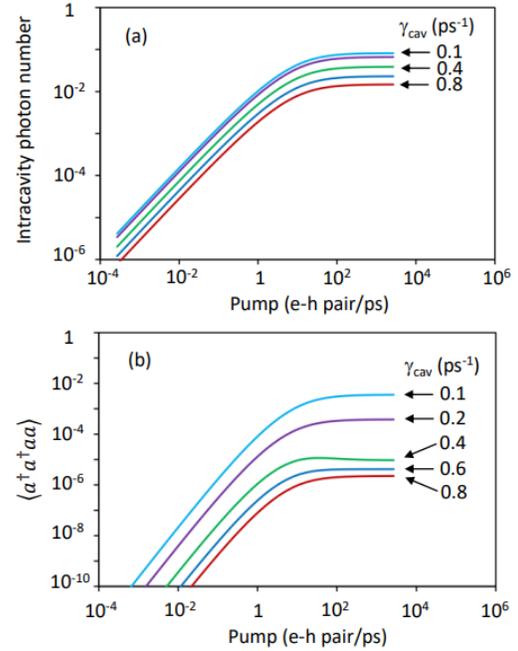

Figure 2: (a) Intracavity photon number $\langle a^\dagger a \rangle$ and (b) measure of multiple photon probability $\langle a^\dagger a^\dagger aa \rangle$ as a function of pump rate. The curves are for different cavity decay rates $\gamma_{cav}$ as indicated.

To further identify the physical mechanisms behind the possible dip in $g^{(2)}(0)$, we list the equations of motion used in the calculations. The derivation starts with the Hamiltonian [25] [26]

$$H = \hbar\nu(a^\dagger a + \frac{1}{2}) + \varepsilon_e c + \varepsilon_h b^\dagger b$$
$$-i\hbar g(b^\dagger c^\dagger a - a^\dagger c\, b), \qquad (2)$$

where $\hbar\nu$ is the photon energy, $\varepsilon_e$ and $\varepsilon_h$ are the QD electron and hole ground-state energies, $c^\dagger$ and $c$ are creation and annihilation operators for electrons, $b^\dagger$ and $b$ are the corresponding operators for holes, and $g$ is the light-matter coupling strength (or vacuum Rabi frequency), expressed as

$$g = \wp\sqrt{\frac{\nu}{\hbar\epsilon_b V}}, \qquad (3)$$

where $\wp$ is the dipole matrix element, $V$ is the optical mode volume and $\epsilon_b$ is the background permittivity. We assume the location of the QD at the cavity mode antinode and ignore electronic structure details such as overlapping of electron and hole envelope functions. For later curves, $g$ is referenced to $\Omega_{R0} = 0.025\ ps^{-1}$, which is the Rabi frequency computed for a $(\lambda/n_B)^3$ cavity with a resonant wavelength $\lambda = 920\ nm$, a background refractive index $n_B = 3.5$, dipole matrix element $\wp = e \times 0.5\ nm$, and $e$ is the electron charge

The photon operators obey commutation relations, while the carrier operators obey anti-commutation relations. Keeping only correlations necessary for the determination of the 2nd order intensity correlation, the cluster expansion gives the equations of motion for the electron, hole, and photon populations:

$$\frac{dn_e}{dt} = -2g\,Re(p) + P(1 - n_e) - \gamma_{nr}n_e$$
$$- \gamma_{nl}n_e n_h, \qquad (4)$$

$$\frac{dn_h}{dt} = -2g\,Re(p) + P(1 - n_h) - \gamma_{nr}n_h$$
$$- \gamma_{nl}n_e n_h, \qquad (5)$$

$$\frac{dn_p}{dt} = 2g\,Re(p) - 2\gamma_c n_p, \qquad (6)$$

$$\frac{dp}{dt} = -[\gamma + \gamma_c + i(\omega - \nu)]p + gn_e n_h$$
$$+ g(n_e + n_h - 1)n_p$$
$$+ g(\delta\langle c^\dagger c a^\dagger a\rangle + \delta\langle b^\dagger b a^\dagger a\rangle). \qquad (7)$$

In the above equations, we introduce $\gamma_{nr}$ the nonradiative carrier loss, $\gamma_{nl}$ the rate of spontaneous emission into free space and all other optical modes, $2\gamma_c$ the photon decay rate in the cavity, and $\gamma$ the dephasing rate. For the excitation, $P$ is the rate of carrier injection, and $(1 - n_\sigma)$ accounts for Pauli blocking. For $g^{(2)}(0)$, we have $\langle a^\dagger a^\dagger aa\rangle = 2n_p^2 + \delta\langle a^\dagger a^\dagger aa\rangle$, where the doublet contribution $\delta\langle a^\dagger a^\dagger aa\rangle$ is given by

$$\frac{d\,\delta\langle a^\dagger a^\dagger aa\rangle}{dt} = -4\gamma_c\,\delta\langle a^\dagger a^\dagger aa\rangle$$

$$+4g\,\text{Re}(\delta\langle b^\dagger c^\dagger a^\dagger aa\rangle). \quad (8)$$

To solve for $\delta\langle a^\dagger a^\dagger aa\rangle$, we need the additional correlations:

$$\frac{d\,\delta\langle b^\dagger c^\dagger a^\dagger aa\rangle}{dt}$$
$$= [-(\gamma + 3\gamma_c) + i(\omega - \nu)]\,\delta\langle b^\dagger c^\dagger a^\dagger aa\rangle$$
$$+ g(n_e + n_h - 1)\delta\langle a^\dagger a^\dagger aa\rangle$$
$$+ 2g(n_h + n_p)\delta\langle c^\dagger c a^\dagger a\rangle$$
$$+ 2g(n_e + n_p)\delta\langle b^\dagger b a^\dagger a\rangle - 2gp^2 \quad (9)$$

$$\frac{d\,\delta\langle c^\dagger c a^\dagger a\rangle}{dt} = -(\gamma_{nr} + 2\gamma_c)\,\delta\langle c^\dagger c a^\dagger a\rangle$$
$$- 2g\text{Re}\big[p(n_e + n_p) + \delta\langle b^\dagger c^\dagger a^\dagger aa\rangle\big] \quad (10)$$

$$\frac{d\,\delta\langle b^\dagger b a^\dagger a\rangle}{dt} = -(\gamma_{nr} + 2\gamma_c)\,\delta\langle b^\dagger b a^\dagger a\rangle$$
$$- 2g\text{Re}\big[p(n_h + n_p) + \delta\langle b^\dagger c^\dagger a^\dagger aa\rangle\big]. \quad (11)$$

Through the process of elimination, we identified $(n_e + n_h - 1)\delta\langle a^\dagger a^\dagger aa\rangle$ (third line of Eq. 9) as the main contribution to the dip in $g^{(2)}(0)$. This term modifies the correlation which describes the reduction of thermal radiation value of 2 and the reduction comes from a decrease in the dephasing rate of which has a dependence on the population inversion. Figure 3 (a) shows the photon correlation $\langle a^\dagger a^\dagger aa\rangle$ versus cavity lifetime for light-matter coupling rate $g = 0.20\,\Omega_{RO}$, with and without the contribution (solid and dashed curves, respectively). Figure 3 (b) shows how the leveling in $\langle a^\dagger a^\dagger aa\rangle$ versus cavity lifetime translates to a dip in $g^{(2)}(0)$.

Figure 4 summarizes the results of a parametric study of $g^{(2)}(0)$ versus photon-production rate. We found three regimes of cavity enhancement. To help understand the distinction, we plot the curves in Figure. 4 (a) for $\gamma_{cav} = 0.3\,ps^{-1}$ (red), $2.2\,ps^{-1}$ (black) and $8\,ps^{-1}$ (blue), illustrating weak-cavity, Purcell enhancement and strong-cavity cases, respectively. In the weak or no cavity situation, $g^{(2)}(0) \approx 0$ for all excitations, but the single-photon flux is low (red curves). With the right balance of spontaneous emission rate and cavity lifetime, there is minimal chance of having a residual photon in the cavity with an excited QD. One gets Purcell enhancement with acceptable compromise in $g^{(2)}(0)$ (black curves). The difference in dashed and solid black curves depicts the role of carrier-photon correlations. With too much cavity enhancement, $g^{(2)}(0)$ suffers from the likelihood of high probability of the presence of intracavity photons (blue curves).

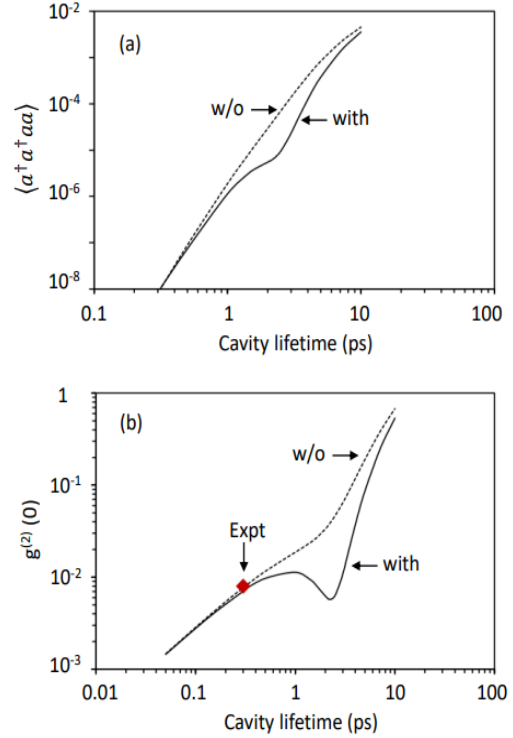

Figure 3: (a) Photon correlation $\langle a^\dagger a^\dagger aa\rangle$ versus cavity lifetime for light-matter coupling strength $g = 0.20\,\Omega_{RO}$, with and without $(n_e + n_h - 1)\delta\langle a^\dagger a^\dagger aa\rangle$ (solid and dashed curves, respectively). (b) Corresponding intensity correlation $g^{(2)}(0)$ versus cavity lifetime. Notations are similar to (a). The red diamond indicates the measured value from [24].

Finally, calculations governed by cQED are performed for the material system described in Figure 1. Figure 5(a) and 5(b)

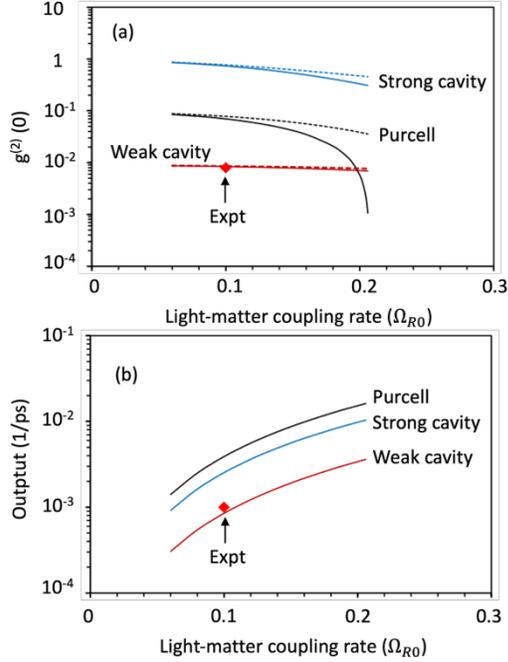

Figure 5: (a) $g^{(2)}(0)$ versus light-matter coupling rate $g$ (relative to $\Omega_{RO} = 0.027\,ps^{-1}$). The solid curves have correlations included at the doublet level. The dashed curves neglects $(n_e + n_h - 1)\delta\langle a^\dagger a^\dagger aa\rangle$ as discussed in the text. (b) Computed photon output rate versus light-matter coupling rate. Curves computed with and without $(n_e + n_h - 1)\delta\langle a^\dagger a^\dagger aa\rangle$ are similar. The switch between Purcell and strong-cavity curves is because of the non-optimal coupling in the latter. The red diamond indicates the measured values from [24].

show calculated results for cavity lifetimes ranging from 0.2 ps (approximating free space) to 10 ps. A pump rate of $10^5$ injected electron-hole pairs per picosecond is used to completely invert the QD carrier population, as in single-photon experiments. As shown in Figure 5(a), all output versus cavity lifetime curves increases with increase $g$.

There appears to be an optimum cavity lifetime after which the output photon rate decreases even though the intracavity photon number continues to grow. Figure 5(b) shows computed dependences of the intensity correlation for the same $g$ values. The $g^{(2)}(0)$ increases with photon generation rate except for when $g = 0.18\,\Omega_{RO}$ and $0.20\,\Omega_{RO}$. The balance between cavity enhancement and spontaneous emission rate results in the desired increase in the single photon production rate while maintaining the high single photon purity.

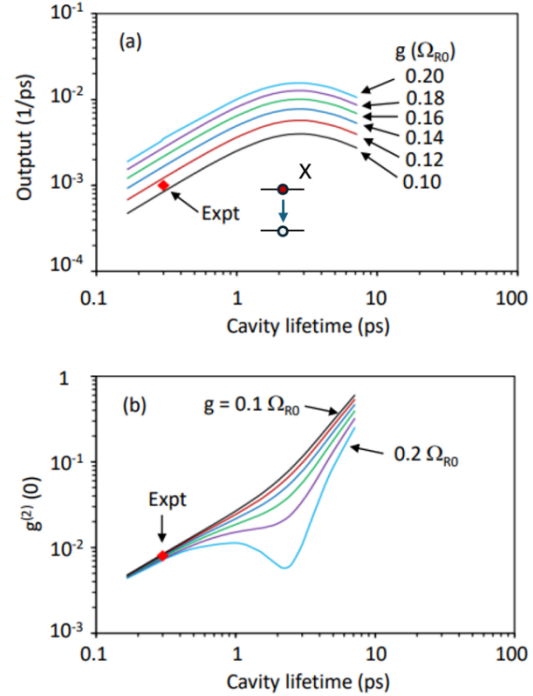

Figure 4: (a) Computed photon output and (b) intensity correlation $g^{(2)}(0)$ as a function of cavity lifetime for various light-matter coupling strength, $g(\Omega_{RO})$. The red diamond indicates the measured values from [27].

Though it is not experimentally possible to adjust cavity lifetime and light-matter coupling strength independently since both parameters depend on the cavity geometry and local field distribution, we could partially control the cavity lifetime without affecting the coupling strength by designing for tapered mirrors. Using photonic crystal cavities can also adjust mode confinement locally while keeping an approximately the same cavity lifetime.

III. CONCLUSION

To summarize, we have performed detailed cQED calculations to investigate the probability of achieving high single photon generation rate while maintaining a high single photon purity. Contrary to common beliefs, we have found that the single photon purity does not degrade monotonically as the generate rate increases in a cavity. With the proper design of the optical cavity to realize the preferred balance between the spontaneous emission rate and cavity lifetime, the single photon purity can be restored with a stronger cavity as the multiphoton probability does not increase as fast as the intracavity photon number. The experimental data in the weak cavity regime anchors well with the calculations. This suggests that the framework we provided here would serve as guidance to the future designs of the emitter-cavity system for fast and pure on-chip single photon generations.

ACKNOWLEDGEMENTS

This work was funded by the NSF Quantum Foundry at UCSB (Grant No. DMR-1906325) and the U.S. Department of Energy under Contract No. DE-AC04-94AL85000 This work was performed, in part, at the Center for Integrated Nanotechnologies, an Office of Science User Facility operated for the U.S. Department of Energy (DOE) Office of Science.